\documentclass{sig-alternate}
\usepackage[english]{babel}
\usepackage{tabularx}
\usepackage{graphicx}
\usepackage{url}
\usepackage{mdwlist}
\usepackage{subcaption}
\usepackage{hyperref}
\usepackage[labelfont=bf,textfont=bf]{caption}
\usepackage{multirow}
\usepackage{array}
\usepackage[linesnumbered,ruled]{algorithm2e}
\SetKwComment{Comment}{$\triangleright$\ }{}
\usepackage{xcolor}

\usepackage[utf8]{inputenc}
\usepackage{amssymb}

\newcommand{\ie}{\emph{i.e., }}
\newcommand{\eg}{\emph{e.g., }}
\newcommand{\etal}{\emph{et al.}}

\newcommand{\wrt}{\emph{w.r.t. }}

\newcommand{\R}{\mathcal{R}}

\begin{document}
	
\CopyrightYear{2016} 
\setcopyright{acmcopyright}
\conferenceinfo{SIGIR '16,}{July 17-21, 2016, Pisa, Italy}
\isbn{978-1-4503-4069-4/16/07}\acmPrice{\$15.00}
\doi{http://dx.doi.org/10.1145/2911451.2911489}

\title{Fast Matrix Factorization for Online Recommendation\\ with Implicit Feedback\thanks{NExT research is supported by the National Research Foundation, Prime Minister's Office, Singapore under its IRC@SG Funding Initiative.}}

\numberofauthors{4}
\author{
	Xiangnan He\qquad
	Hanwang Zhang\qquad
	Min-Yen Kan\qquad
	Tat-Seng Chua\\
	\affaddr{School of Computing, National University of Singapore}\\
	\email{\{xiangnan, hanwang, kanmy, chuats\}@comp.nus.edu.sg}
}
\maketitle
\begin{abstract}	



This paper contributes improvements on both the effectiveness and efficiency of \textit{Matrix Factorization} (MF) methods for implicit feedback. We highlight two critical issues of existing works. First, due to the large space of unobserved feedback, most existing works resort to assign a uniform weight to the missing data to reduce computational complexity. However, such a uniform assumption is invalid in real-world settings. Second, most methods are also designed in an offline setting and fail to keep up with the dynamic nature of online data. 

We address the above two issues in learning MF models from implicit feedback. We first propose to weight the missing data based on item popularity, which is more effective and flexible than the uniform-weight assumption. However, such a non-uniform weighting poses efficiency challenge in learning the model.
To address this, we specifically design a new learning algorithm based on the \textbf{e}lement-wise \textbf{A}lternating \textbf{L}east \textbf{S}quares (eALS) technique, for efficiently optimizing a MF model with variably-weighted missing data. We exploit this efficiency to then seamlessly devise an incremental update strategy that instantly refreshes a MF model given new feedback. Through comprehensive experiments on two public datasets in both offline and online protocols, we show that our eALS method consistently outperforms state-of-the-art implicit MF methods. Our implementation is available at \url{https://github.com/hexiangnan/sigir16-eals}.

\end{abstract}


\vspace{-5pt}
\vspace{-5pt}

\section{Introduction}
\label{sec:introduction}

User personalization has become prevalent in modern recommender system. It helps to capture users' individualized preferences and has been shown to increase both satisfaction for users and revenue for content providers. 
Among its various methods, matrix factorization (MF) is the most popular and effective technique that characterizes users and items by vectors of latent factors~\cite{koren2011advances,DCF:2016}.
Early work on MF algorithms for recommendation \cite{Koren:timeSVD, Rendle:2008} have largely focused on explicit feedback, where users' ratings that directly reflect their preference on items are provided. These works formulated recommendation as a rating prediction problem for which the large volume of unobserved ratings (\ie missing data) are assumed to be extraneous for modeling user preference~\cite{Hu:2008}. This greatly reduces the modeling workload, and many sophisticated methods have been devised, such as SVD++~\cite{koren2011advances} and timeSVD~\cite{Koren:timeSVD}.

However, explicit ratings are not always available in many applications; more often, users interact with items through \textbf{implicit feedback}, \eg users' video viewing and product purchase history. 
Compared to explicit ratings, implicit feedback is easier to collect for content providers, but more challenging to utilize due to the natural scarcity of negative feedback. 
It has been shown that modeling only the observed, positive feedback results in biased representations in user profiles~\cite{devooght2015dynamic, Hu:2008}; \eg Marlin \etal~\cite{Marlin07MAR} finds that users listen to music they expect to like and avoid the genres they dislike, leading to a severe bias in the observed data. 

To solve the problem of lacking negative feedback (also known as the one-class problem \cite{Pan:2008}), a popular solution is to model all the missing data as negative feedback \cite{Hu:2008}. However, this adversely degrades the learning efficiency due to the full consideration of both observed and missing data. More importantly, the low efficiency makes it even more difficult to deploy implicit MF method \textbf{online} \cite{MLGoogle:2014}. In practical recommender systems where new users, items and interactions are continuously streaming in, it is crucial to refresh the underlying model in real-time to best serve users. 
In this work, we concern the above two challenging problems of the MF method ---  implicit feedback and online learning. We note that we are not the first to consider both aspects for MF, as a recent work by Devooght \etal~\cite{devooght2015dynamic} has proposed an efficient implicit MF method for learning with dynamic data. However, we argue that Devooght's method~\cite{devooght2015dynamic} models missing data in an unrealistic, suboptimal way.
Specifically, it assigns a uniform weight to the missing data, assuming that the missing entries are equally likely to be negative feedback. However, such an assumption limits model's fidelity and flexibility for real applications. For example, content providers usually know which items have been frequently featured to users but seldom clicked; such items are more likely to be true negative assessments and should be weighted higher than others. In addition, Devooght's method learns parameters through gradient descent, requiring an expensive line search to determine the best learning rate at each step. 

We propose a new MF method aimed at learning from implicit feedback effectively while satisfying the requirement of online learning. We develop a new learning algorithm that efficiently optimizes the implicit MF model without imposing a uniform-weight restriction on missing data. In particular, we assign the weight of missing data based on the popularity of items, which is arguably more effective than the previous methods \cite{devooght2015dynamic,Hu:2008,Pilaszy:2010,Steck:2010,Volkovs:2015} that are limited by the uniformity assumption. 
Our eALS algorithm is fast in accounting for missing data --- analytically $K$ times faster than ALS~\cite{Hu:2008} where $K$ denotes number of latent factors --- the same time complexity with the recent dynamic MF solution \cite{devooght2015dynamic}. This level of efficiency makes eALS suitable for learning online, for which we develop an incremental update strategy that instantly refreshes model parameters given new incoming data. Another key advantage of eALS is that it works without learning rate, bypassing the well-known difficulty for tuning gradient descent methods such as \cite{devooght2015dynamic} and \textit{Stochastic Gradient Descent} (SGD) \cite{Rendle:2009:BPR}.

We summarize our key contributions as follows. \vspace{-5pt}
\begin{enumerate}
\item We propose an item popularity-aware weighting scheme on the full missing data that effectively tailors the MF model for learning from implicit feedback. \vspace{-5pt}
\item We develop a new algorithm for learning model parameters efficiently and devise an incremental update strategy to support real-time online learning. \vspace{-5pt}
\item We conduct extensive experiments with both offline and online protocols on two real-world datasets, showing that our method consistently outperforms state-of-the-art implicit MF methods.
\end{enumerate}

\section{Related Work}
\label{sec:related}


Handling missing data is obligatory for learning from implicit data due to the lack of negative feedback. To this end, two strategies have been proposed --- \textbf{sample based learning} \cite{Pan:2008,Rendle:2009:BPR} that samples negative instances from missing data, or \textbf{whole-data based learning}~\cite{Hu:2008,Steck:2010} that treats all missing data as negative. 
Both methods have pros and cons: sample-based methods are more efficient by reducing negative examples in training, but risk decreasing the model's predictiveness; whole-based methods model the full data with a potentially higher coverage, but inefficiency can be an issue. To retain model's fidelity, we persist in the whole-data based learning, developing a fast ALS-based algorithm to resolve the inefficiency issue.

For existing whole-data based methods \cite{devooght2015dynamic,Hu:2008,Pilaszy:2010,Steck:2010,Volkovs:2015}, one major limitation is in the uniform weighting on missing entries, which favors algorithm's efficiency but limits model's flexibility and extensibility. 
The only works that have considered non-uniform weighting are from Pan \etal~\cite{Pan:2009,Pan:2008}; however their cubic time complexity \wrt $K$ makes it unsuitable to run on large-scale data~\cite{Pilaszy:2010}, where a large number of factors needs to be considered to gain improved performance \cite{Volkovs:2015}.  



To optimize MF, various learners have been investigated, including SGD \cite{Koren:timeSVD,Rendle:2009:BPR}, \textit{Coordinate Descent} (CD) \cite{devooght2015dynamic,DCF:2016}, and \textit{Markov Chain Monto Carlo} (MCMC) \cite{libfm}. SGD is the most popular one owing to the ease of derivation,
however, it is unsuitable for whole-data based MF \cite{Hu:2008} due to the large amount of training instances (the full user--item interaction matrix is considered). ALS can be seen as an instantiation of CD and
has been widely used to solve the whole-based MF \cite{Hu:2008,Pan:2009,Pan:2008,Steck:2010}; however, its inefficiency is the main obstacle for practical use \cite{Pilaszy:2010,Volkovs:2015}. To resolve this, \cite{Pilaszy:2010} describes an approximate solution to ALS.
Recently, \cite{devooght2015dynamic} employs the \textit{Randomized block Coordinate Descent} (RCD) learner~\cite{richtarik2014iteration}, reducing the complexity 
and applying it to a dynamic scenario.  Similarly, \cite{Volkovs:2015} enriches the implicit feedback matrix with neighbor-based similarly, followed by applying unweighted SVD.
Distinct from previous works, we propose an efficient element-wise ALS solution for the whole-data based MF with non-uniform missing data, which has never been studied before. 

Another important aspect for practical recommender system lies in handling the dynamic nature of incoming data, for which timeliness is a key consideration. As it is prohibitive to retrain the full model online, various works have developed incremental learning strategies for neighbor-based \cite{Huang:2015}, graph-based~\cite{He:2015}, probabilistic~\cite{Das:2007} and MF~\cite{devooght2015dynamic,diaz2012real,ling2012online,Rendle:2008} methods.
For MF, different learners have been studied for online updating, including SGD \cite{diaz2012real,Rendle:2008}, RCD \cite{devooght2015dynamic} and dual-averaging \cite{ling2012online}. To our knowledge, this work is the first attempt to exploit the ALS technique for online learning. 



 
\section{Preliminaries}
\label{sec:mf}
We first introduce the whole-data based MF method for learning from implicit data, highlighting the inefficiency issue of the conventional ALS solution \cite{Hu:2008,Pan:2008}. Then we describe eALS, an element-wise ALS learner~\cite{libfm} that can reduce the time complexity to linearity \wrt number of factors. Although the learner is generic in optimizing MF with all kinds of weighting strategies, the form introduced is costly in accounting for all missing data and is thus unrealistic for practical use.  This defect motivates us to further develop eALS to make it suitable for learning from implicit feedback (details in Section~\ref{ss:efficient_eals}).

\subsection{MF Method for Implicit Feedback}
We start by introducing some basic notation. 
For a user--item interaction matrix $\textbf{R}\in \mathbb{R}^{M\times N}$, $M$ and $N$ denote the number of users and items, respectively; $\R$ denotes the set of user--item pairs whose values are non-zero. We reserve the index $u$ to denote a user and $i$ to denote an item. Vector $\textbf{p}_u$ denotes the latent feature vector for $u$, and set $\R_u$ denotes the set of items that are interacted by $u$; similar notations for $\textbf{q}_i$ and $\R_i$. 
Matrices $\textbf{P}\in \mathbb{R}^{M\times K}$ and $\textbf{Q}\in \mathbb{R}^{N\times K}$ denote the latent factor matrix for users and items. 

Matrix factorization maps both users and items into a joint latent feature space of $K$ dimension such that interactions are modeled as inner products in that space. Mathematically, each entry $r_{ui}$ of $\textbf{R}$ is estimated as:
\begin{equation}
\label{eq:mf}
\hat{r}_{ui} = <\textbf{p}_u, \textbf{q}_i> = \textbf{p}_u^T \textbf{q}_i.
\end{equation}
The item recommendation problem is formulated as estimating the scoring function $\hat{r}_{ui}$, which is used to rank items.
Note that this basic model subsumes the biased MF~\cite{Koren:timeSVD}, commonly used in modeling explicit ratings:
\begin{equation*}
	\hat{r}_{ui} = b_u + b_i + <\textbf{p}^B_u, \textbf{q}^B_i>,
\end{equation*}
where $b_u$ ($b_i$) captures the bias of user $u$ (item $i$) in giving (receiving) ratings. To recover it, set $\textbf{p}_u \leftarrow [\textbf{p}^B_u, b_u, 1]$ and $\textbf{q}_i \leftarrow [\textbf{q}^B_i, 1, b_i]$. As such, we adopt the basic MF model to make notations simple and also to enable a fair comparison with baselines~\cite{devooght2015dynamic,Hu:2008} that also complied with the basic model. 

To learn model parameters, Hu \etal~\cite{Hu:2008} introduced a weighted regression function, which associates a confidence to each prediction in the implicit feedback matrix $\textbf{R}$:
\begin{equation}
\label{eq:J}
J = \sum_{u=1}^M \sum_{i=1}^N w_{ui}(r_{ui} - \hat{r}_{ui})^2 + \lambda(\sum_{u=1}^M ||\textbf{p}_u||^2 + \sum_{i=1}^N ||\textbf{q}_i||^2),
\end{equation}
where $w_{ui}$ denotes the weight of entry $r_{ui}$ and we use $\textbf{W}=[w_{ui}]_{M\times N}$ to represent the weight matrix.
$\lambda$ controls the strength of regularization, which is usually an $L_2$ norm to prevent overfitting. Note that in implicit feedback learning, missing entries are usually assigned to a zero $r_{ui}$ value but non-zero $w_{ui}$  weight, both crucial to performance. 

\subsection{Optimization by ALS}
Alternating Least Square (ALS) is a popular approach to optimize regression models such as MF and graph regularization~\cite{He:SIGIR14}. It works by iteratively optimizing one parameter, while leaving the others fixed. The prerequisite of ALS is that the optimization sub-problem can be analytically solved. Here, we describe how Hu's work \cite{Hu:2008} solves this problem.

First, minimizing $J$ with respect to user latent vector $\textbf{p}_u$ is equivalent to minimizing:
\begin{equation*}
	\begin{aligned}
		\label{eq:loss_u}
		J_u = ||\textbf{W}^u(\textbf{r}_u - \textbf{Q}\textbf{p}_u) ||^2 + \lambda ||\textbf{p}_u||^2,
	\end{aligned}
\end{equation*}
where $\textbf{W}^u$ is a $N\times N$ diagonal matrix with $W^u_{ii} = w_{ui}$. The minimum is where the first-order derivative is 0:
\begin{equation}
\begin{aligned}
\label{eq:p_u}
&\frac{\partial J_u}{\partial \textbf{p}_u} = 2\textbf{Q}^T \textbf{W}^u\textbf{Q} \textbf{p}_u - 2\textbf{Q}^T \textbf{W}^u \textbf{r}_u + 2\lambda\textbf{p}_u = 0 \\
\Rightarrow & \quad\textbf{p}_u = (\textbf{Q}^T \textbf{W}^u \textbf{Q} + \lambda\textbf{I})^{-1} \textbf{Q}^T\textbf{W}^u \textbf{r}_u,
\end{aligned}
\end{equation}
where $\textbf{I}$ denotes the identity matrix.
This analytical solution is also known as the \textit{ridge regression} \cite{Pilaszy:2010}. 
Following the same process, we can get the solution for $\textbf{q}_i$.

\subsubsection{Efficiency Issue with ALS}
As we can see, in order to update a latent vector, inverting a $K\times K$ matrix is inevitable. Matrix inversion is an expensive operation, usually assumed $O(K^3)$ in time complexity~\cite{Hu:2008}. As such, updating one user latent vector takes time $O(K^3 + N K^2)$. Thus, the overall time complexity of one iteration that updates all model parameters once is $O((M+N)K^3 + M N K^2)$. Clearly, this high complexity makes the algorithm impractical to run on large-scale data, where there can be millions of users and items and billions of interactions.

\textbf{Speed-up with Uniform Weighting}. 
To reduce the high time complexity, Hu \etal~\cite{Hu:2008} applied a uniform weight to missing entries; {\it i.e.}, assuming that all zero entries in $\textbf{R}$ have a same weight $w_0$. Through this simplification, they can speed up the computation with memoization:
\begin{equation}
\label{eq:QWQ}
\textbf{Q}^T \textbf{W}^u \textbf{Q} = w_0\textbf{Q}^T \textbf{Q} + \textbf{Q}^T (\textbf{W}^u - \textbf{W}^0) \textbf{Q},
\end{equation}
where $\textbf{W}^0$ is a diagonal matrix that each diagonal element is $w_0$. As $\textbf{Q}^T \textbf{Q}$ is independent of $u$, it can be pre-computed for updating all user latent vectors. Considering the fact that $\textbf{W}^u - \textbf{W}^0$ only has $|\R_u|$ non-zero entries, we can compute Eq.~(\ref{eq:QWQ}) in $O(|\R_u| K^2)$ time. Thus, the time complexity of ALS is reduced to $O((M+N)K^3 + |\R|K^2)$. 

Even so, we argue that the $O((M+N)K^3)$ term can be a major cost when $(M+N)K\ge |\R|$. In addition, the $O(|\R|K^2)$ part is still much higher than in SGD \cite{Rendle:2009:BPR}, which only requires $O(|\R|K)$ time. 
As a result, even with the acceleration, ALS is still prohibitive for running on large data, where large $K$ is crucial as it can lead to better generalizability and thus better prediction performance.
Moreover, the uniform weighting assumption is usually invalid in real applications and adversely degrades model's predictiveness. 
This thus motivates us to design an efficient implicit MF method not subject to uniform-weights. 


\subsection{Generic Element-wise ALS Learner}
\label{ss:generic_eals}
The bottleneck of the previous ALS solution lies in the matrix inversion operation, which is due to the design that updates the latent vector for a user (item) as a whole. As such, it is natural to optimize parameters at the element level --- optimizing each coordinate of the latent vector, while leaving the others fixed \cite{libfm}.
To achieve this, we first get the derivative of objective function Eq.~(\ref{eq:J}) with respect to $p_{uf}$:
\vspace{-2mm}
\begin{equation*}
	\frac{\partial J}{\partial p_{uf}} = -2\sum_{i=1}^N (r_{ui} - \hat{r}_{ui}^f)w_{ui}q_{if} + 2p_{uf}\sum_{i=1}^N w_{ui}q_{if}^2 + 2\lambda p_{uf},
\end{equation*}
where $\hat{r}_{ui}^f = \hat{r}_{ui} - p_{uf}q_{if}$, \ie the prediction without the component of latent factor $f$. By setting this derivative to 0, we obtain the solution of $p_{uf}$:
\begin{equation}
\label{eq:p_uf0}
p_{uf} = \frac{\sum_{i=1}^N (r_{ui} - \hat{r}_{ui}^f)w_{ui}q_{if}}{\sum_{i=1}^N w_{ui}q_{if}^2 + \lambda}.
\end{equation}

Similarly, we can get the solver for an item latent factor:
\begin{equation}
\label{eq:q_if0}
q_{if} = \frac{\sum_{u=1}^M (r_{ui} - \hat{r}_{ui}^f)w_{ui}p_{uf}}{\sum_{i=1}^M w_{ui}p_{uf}^2 + \lambda}.
\end{equation}

Given the closed-form solution that optimizes one parameter with other fixed, the algorithm iteratively executes it for all model parameters until a joint optimum is reached. Due to the non-convexity of the objective function, critical points where gradients vanish can be local minima. 

\textbf{Time Complexity}.
As can be seen, by performing optimization at the element level, the expensive matrix inversion can be avoided. A raw implementation takes $O(M N K^2)$ time for one iteration, directly speeding up ALS by eliminating the $O(K^3)$ term. 
Moreover, by pre-computing $\hat{r}_{ui}$~\cite{libfm}, we can calculate  $\hat{r}_{ui}^f$ in $O(1)$ time rather than $O(K)$.
As such, the complexity can be further reduced to $O(M N K)$, which is the same magnitude with evaluating all the user--item predictions.

\section{Our Implicit MF Method}
\label{sec:method}

We first propose an item-oriented weighting scheme on the missing data, and follow with a popularity-aware weighting strategy, which is arguably more effective than the uniform weighting for the recommendation task.
Then, we develop a fast eALS algorithm to optimize the objective function that significantly reduces learning complexity comparing with the conventional ALS~\cite{Hu:2008} and generic element-wise ALS learner~\cite{libfm}. 
Lastly, we discuss how to adjust the learning algorithm for real-time online learning.


\subsection{Item-Oriented Weighting on Missing Data}
\label{ss:popularity}

Due to the large space of items, the missing entries for a user are a mixture of negative and unknown feedback. In specifying the weight $w_{ui}$ of missing entries, it is desired to assign a higher weight to the negative feedback. However, it is a well-known difficulty to differentiate the two cases. In addition, as the interaction matrix $\textbf{R}$ is usually large and sparse, it will be too consuming to store each zero entry an individualized weight. To this end, existing works \cite{devooght2015dynamic,Hu:2008,Pilaszy:2010,Steck:2010,Volkovs:2015} have applied a simple uniform weight on missing entries, which are, however, suboptimal and non-extendable for real applications. 

Considering the ease of content providers in accessing negative information of the item side (\eg which items have been promoted to users but receive little interaction), we believe it is more realistic to weight missing data based on some item property. To capture this, we devise a more fine-grained objective function as follows:
\begin{equation}
	\small
	\begin{aligned}
		\label{eq:L}
		L &= \sum_{(u,i)\in \R} w_{ui}(r_{ui} - \hat{r}_{ui})^2 
		+ \sum_{u=1}^M \sum_{i\notin \R_u} c_i\hat{r}_{ui}^2 \\
		&+ \lambda(\sum_{u=1}^M ||\textbf{p}_u||^2 + \sum_{i=1}^N ||\textbf{q}_i||^2),
	\end{aligned}
\end{equation}
where $c_i$ denotes the confidence that item $i$ missed by users is a true negative assessment, which can serve as a means to encode domain knowledge from practitioners. It is clear that the first term denotes the prediction error of the observed entries, which has been widely adopted in modeling explicit ratings~\cite{koren2011advances,Rendle:2008}. The second term accounts for the missing data, which acts as the role of negative instances and is crucial for recommendation from implicit feedback \cite{Hu:2008,Rendle:2009:BPR}.
Next, we present a domain-independent strategy to determine $c_i$ by leveraging a ubiquitous feature of modern Web 2.0 systems. 




\subsubsection{Popularity-aware Weighting Strategy}
Existing visual interfaces of many Web 2.0 systems showcase popular items in their recommendations. All other factors being equal, popular items are more likely to be known by users in general~\cite{He:SIGIR14}, and thus it is reasonable to think that a miss on a popular item is more probable to be truly irrelevant (as opposed to unknown) to the user.
To account for this effect, we parametrize $c_i$ based on item's popularity:
\begin{equation}
\small
c_i = c_0 \frac{f_i^\alpha}{\sum_{j=1}^N f_j^\alpha},
\end{equation}
where $f_i$ denotes the popularity of item $i$, given by its frequency in the implicit feedback data: $|\R_i|/\sum_{j=1}^N{|\R_j|}$, and $c_0$ determines the overall weight of missing data. Exponent $\alpha$ controls the significance level of popular items over unpopular ones --- when $\alpha > 1$ the weights of popular items are promoted to strengthen the difference against unpopular ones; while setting $\alpha$ within the lower range of $(0,1)$ suppresses the weight of popular items and has a smoothing effect. We empirically find $\alpha=0.5$ usually leads to good results.
Note that the uniform weighting is a special case by setting $\alpha$ to 0 with $w_0=c_0/N$. 

\textbf{Relationship to Negative Sampling}.
Our proposed popularity-aware weighting strategy has the same intuition with Rendle's popularity-based oversampling \cite{Rendle:2014:IPL} for learning BPR, which basically samples popular items as negative feedback with a higher probability. However, \cite{Rendle:2014:IPL} empirically shows the oversampling method underperforms the basic uniform sampler.
We suspect the reason comes from the SGD learner, which will result in more gradient steps on popular items, due to oversampling. As a result, popular items may be over-trained locally at the expense of less popular items which would then be under-trained. 
To resolve this, tricks like subsampling frequent items \cite{mikolov2013distributed} and adaptive learning rates like Adagrad \cite{Adagrad} have been adopted in other domains. 
As the focus of this paper is on whole-data based implicit MF, we do not further explore the details of SGD. It is worth pointing out that our proposed eALS learner avoids these learning issues by an exact optimization on each model parameter. 

\subsection{Fast eALS Learning Algorithm}
\label{ss:efficient_eals}
We can speed up learning by avoiding the massive repeated computations introduced by the weighted missing data.  We detail the derivation process for $p_{uf}$; where the counterpart for $q_{if}$ is achieved likewise.

First, we rewrite the $p_{uf}$ update rule Eq.~(\ref{eq:p_uf0}) by separating the observed data part:
\begin{equation*}
\begin{aligned}
p_{uf} &= \frac{\sum_{i\in \R_u} (r_{ui} - \hat{r}_{ui}^f) w_{ui} q_{if} - \sum_{i\notin \R_u} \hat{r}_{ui}^f c_i q_{if}}{\sum_{i\in \R_u} w_{ui}q_{if}^2 + \sum_{i\notin \R_u} c_i q_{if}^2 + \lambda }.
\end{aligned}
\end{equation*}

Clearly, the computational bottleneck lies in the summation over missing data portion, which requires a traversal of the whole negative space. We first focus on the numerator:
\begin{equation}
\begin{aligned}
\label{eq:numer}
\sum_{i\notin \R_u} \hat{r}_{ui}^f c_{i}q_{if} 
&= \sum_{i=1}^N c_i q_{if}\sum_{k\ne f} p_{uk}q_{ik} - \sum_{i\in \R_u} \hat{r}_{ui}^f c_i q_{if} \\
&= \sum_{k\ne f} p_{uk} \sum_{i=1}^N c_i q_{if} q_{ik} - \sum_{i\in \R_u} \hat{r}_{ui}^f c_i q_{if}.
\end{aligned}
\end{equation}
By this reformulation, we can see that the major computation --- the $\sum_{i=1}^N c_i q_{if} q_{ik}$ term that iterates over all items --- is independent of $u$. However, a na\"{\i}ve implementation repeatedly computes it unnecessarily, when updating the latent factors for different users. Clearly, we can achieve a significant speed-up by memoizing it.

We define the $\textbf{S}^q$ cache as $\textbf{S}^q = \sum_{i=1}^N  c_i\textbf{q}_i \textbf{q}_i^T$, which 
can be pre-computed and used in updating the latent factors for all users. Then, Eq.~(\ref{eq:numer}) can be evaluated as:
\begin{equation}
\sum_{i\notin \R_u} \hat{r}_{ui}^f c_{i}q_{if} = \sum_{k\ne f} p_{uk} s^q_{fk} - \sum_{i\in \R_u} \hat{r}_{ui}^f c_i q_{if},
\end{equation}
which can be done in $O(K + |\R_u|)$ time. 

Similarly, we can apply the cache to speed up the calculation of denominator:
\begin{equation}
\sum_{i\notin \R_u} c_i q_{if}^2 = \sum_{i=1}^N c_i q_{if}^2 - \sum_{i\in \R_u} c_i q_{if}^2 = s^q_{ff} - \sum_{i\in \R_u} c_i q_{if}^2.
\end{equation}

To summarize the above memoization strategy, we give the update rule for $p_{uf}$ with the use of $\textbf{S}^q$ cache:
\begin{equation}
\label{eq:p_uf}
p_{uf} = \frac{\sum_{i\in \R_u} [w_{ui}r_{ui} - (w_{ui} - c_i)\hat{r}_{ui}^f] q_{if} - \sum_{k\neq f} p_{uk} s^q_{kf}}{\sum_{i\in \R_u} (w_{ui} - c_i)q_{if}^2 + s^q_{ff} + \lambda }.
\end{equation}
Similarly, we can derive the update rule for $q_{if}$:
\begin{equation}
\label{eq:q_if}
q_{if} = \frac{\sum_{u\in \R_i} [w_{ui}r_{ui} - (w_{ui} - c_i)\hat{r}_{ui}^f] p_{uf} - c_i\sum_{k\neq f} q_{ik} s^p_{kf}}{\sum_{u\in \R_i} (w_{ui} - c_i)p_{uf}^2 + c_i s^p_{ff} + \lambda },
\end{equation}
where $s^p_{kf}$ denotes the $(k,f)^{th}$ element of the $\textbf{S}^p$ cache, defined as $\textbf{S}^p = \textbf{P}^T \textbf{P}$.

Algorithm~\ref{alg:eALS2} summarizes the accelerated algorithm for our element-wise ALS learner, or \textbf{eALS}. For convergence, one can either monitor the value of objective function on training set or check the prediction performance on a hold-out validation data. 


\begin{algorithm}[t]
	\caption{Fast eALS Learning algorithm.}
	\label{alg:eALS2}
	\KwIn{$\textbf{R}$, $K$, $\lambda$, $\textbf{W}$ and item confidence vector $\textbf{c}$;}
	\KwOut{Latent feature matrix $\textbf{P}$ and $\textbf{Q}$;}
	Randomly initialize $\textbf{P}$ and $\textbf{Q}$ \;
	\textbf{for} $(u,i)$$\in \R$ \textbf{do}\quad $\hat{r}_{ui}\leftarrow$ Eq.~(\ref{eq:mf})  \Comment*[r]{$O(|\R| K)$}
	\While{Stopping criteria is not met} {
		\tcp{Update user factors}
		$\textbf{S}^q = \sum_{i=1}^N  c_i\textbf{q}_i \textbf{q}_i^T$ \Comment*[r]{$O(M K^2)$}
		\For(\Comment*[f]{$O(M K^2 + |\R|K)$}){$u \gets 1$ \textbf{\upshape{to}} $M$} {
			\For{$f \gets 1$ \textbf{\upshape{to}} $K$} {
				\textbf{for} $i\in \R_u$ \textbf{do}  $\hat{r}_{ui}^f\leftarrow \hat{r}_{ui} - p_{uf}q_{if}$\;
				$p_{uf}\leftarrow$ Eq.~(\ref{eq:p_uf})  \Comment*[r]{$O(K + |\R_u|)$}
				\textbf{for} $i\in \R_u$ \textbf{do} $\hat{r}_{ui}\leftarrow \hat{r}_{ui}^f + p_{uf}q_{if}$\;
			}
		}
		\tcp{Update item factors}
		$\textbf{S}^p \gets \textbf{P}^T \textbf{P}$ \Comment*[r]{$O(N K^2)$}
		\For(\Comment*[f]{$O(N K^2 + |\R|K)$}){$i \gets 1$ \textbf{\upshape{to}} $N$} {
			\For{$f \gets 1$ \textbf{\upshape{to}} $K$} {
				\textbf{for} $u\in \R_i$ \textbf{do}  $\hat{r}_{ui}^f\leftarrow \hat{r}_{ui} - p_{uf}q_{if}$\;
				$q_{if}\leftarrow$ Eq.~(\ref{eq:q_if})  \Comment*[r]{$O(K + |\R_i|)$}
				\textbf{for} $u\in \R_i$ \textbf{do} $\hat{r}_{ui}\leftarrow \hat{r}_{ui}^f + p_{uf}q_{if}$\;
			}
		}
	}
	\Return{$\textbf{P}$ and $\textbf{Q}$}
\end{algorithm}

\subsubsection{Discussion}
\textbf{Time Complexity}. 
In Algorithm \ref{alg:eALS2}, updating a user latent factor takes $O(K + |\R_u|)$ time. Thus, one eALS iteration takes $O((M+N)K^2 + |R|K)$ time. 
Table~\ref{tab:complexity} summarizes the time complexity (of one iteration or epoch) of other MF algorithms that are designed for implicit feedback. 

\begin{table}[t]
\begin{center}
\caption{\textbf{Time complexity of implicit MF methods.}}
\vspace{-8pt}
\label{tab:complexity}
    \begin{tabular}{ | l | c | c | c | c | }
    \hline
    \textbf{Method} & \textbf{Time Complexity} \\ \hline
    ALS~(Hu \etal \cite{Hu:2008})	&  $O((M+N)K^3 + |\R| K^2)$ \\ \hline
    BPR~(Rendle \etal \cite{Rendle:2009:BPR}) & $O(|\R|K)$ \\ \hline
    IALS1~(Pil\'{a}szy \etal \cite{Pilaszy:2010}) & $O(K^3 + (M+N)K^2 + |\R|K)$ \\ \hline
    ii-SVD~(Volkovs \etal \cite{Volkovs:2015}) & $O((M+N)K^2 + MN\log K)$ \\ \hline
    RCD~(Devooght \etal \cite{devooght2015dynamic})	& $O((M+N)K^2 + |\R|K)$ \\ \hline
    eALS (Algorithm \ref{alg:eALS2}) & $O((M+N)K^2 + |\R|K)$ \\ \hline
    \end{tabular}
\end{center}
\vspace{-7pt}
    \scriptsize{$|\R|$ denotes the number of non-zeros in user--item matrix $\textbf{R}$.}
    \vspace{-10pt}
\end{table}

Comparing with the vector-wise ALS \cite{Hu:2008,Pan:2008}, our element-wise ALS learner is $K$ times faster. 
In addition, our proposed eALS has the same time complexity with RCD~\cite{devooght2015dynamic}, being faster than ii-SVD \cite{Volkovs:2015}, another recent solution. RCD is a state-of-the-art learner for whole-data based MF, which performs a gradient descent step on a randomly chosen latent vector. Since it requires a good learning rate, the work \cite{devooght2015dynamic} adaptively determines it by a line search in each gradient step, which essentially chooses the
learning rate that leads to the steepest descent among pre-defined candidates.
A major advantage of eALS has over RCD is that it avoids the need for a learning rate by an exact optimization in each parameter update, arguably more effective and easier to use than RCD. The most efficient algorithm is BPR, which applies the SGD learner on sampled, partial missing data only.

\textbf{Computing the Objective Function}. Evaluating the objective function is important to check the convergence of iterations and also to verify the correctness of implementation. A direct calculation takes $O(MNK)$ time, requiring a full estimation on the $\textbf{R}$ matrix.
Fortunately, with the item-oriented weighting, we can similarly exploit the sparseness of $\textbf{R}$ for acceleration. To achieve this, we reformulate the loss of the missing data part that causes the major cost:
\vspace{-5pt}
\begin{equation}
\small
\sum_{u=1}^M \sum_{i\notin \R_u} c_i \hat{r}_{ui}^2 
= \sum_{u=1}^M \textbf{p}_u^T \textbf{S}^q \textbf{p}_u - \sum_{(u,i)\in\R } c_i \hat{r}_{ui}^2.
\end{equation}
By reusing $\textbf{S}^q$ and the prediction cache $\hat{r}_{ui}$, we can calculate the objective function in $O(|\R|+MK^2)$ time, much faster than with direct calculation.

\textbf{Parallel Learning}. The iterations of eALS can be easily parallelized. First, computing the $\textbf{S}$ caches (line 4 and 12) is the standard matrix multiplication operation, for which modern matrix toolkits provide very efficient and parallelized implementation. Second, in updating the latent vectors for different users (line 5-11), the shared parameters are either independent with each other (\ie $\hat{r}_{ui}$) or remaining unchanged (\ie $\textbf{S}^q$). 
This nice property means that an exact parallel solution can be obtained by separating the updates by users; that is, letting different workers update the model parameters for disjoint sets of users. The same parallelism can also be achieved in updating item latent vectors. 

This is an advantage over the commonly-used SGD learner, which is a stochastic method that updates model parameters given a training instance. In SGD, different gradient steps can influence with each other and there is no exact way to separate the updates for workers. Thus, sophisticated strategies are required to control the possible losses introduced by parallelization~\cite{gemulla2011large}. Our proposed eALS solution optimizes by coordinate descent where in each step a dedicated parameter is updated, making the algorithm embarrassingly parallel without any approximate loss. 

\subsection{Online Update}
In practice, after a recommender model is trained offline on historical data, 
it will be used online and will need to adapt to best serve users. Here, we consider the online learning scenario that refreshes model parameters given a new user--item interaction.

\textbf{Incremental Updating}. Let $\hat{\textbf{P}}$ and $\hat{\textbf{Q}}$ denote the model parameters learnt from offline training, and $(u, i)$ denotes the new interaction streamed in. To approximate the model parameters in accounting for the new interaction, we perform optimization steps for $\textbf{p}_u$ and $\textbf{q}_i$ only. The underlying assumption is that the new interaction should not change $\hat{\textbf{P}}$ and $\hat{\textbf{Q}}$ too much from a global perspective, while it should change the local features for $u$ and $i$ significantly.
Particularly, when $u$ is a new user, executing the local updates will force $\textbf{p}_u$ close to $\textbf{q}_i$, which meets the expectation of latent factor model. The new item case is similar.

Algorithm~\ref{alg:online} summarizes the incremental learning strategy for eALS.
For the stopping criteria, our empirical study shows that one iteration is usually sufficient to get good results. Moreover, it is important to note that after updating a latent vector, we need to update the $\textbf{S}$ cache accordingly. 

\textbf{Weight of New Interactions}. In an online system, new interactions are more reflective of a user's short-term interest. Comparing to the historical interactions used in offline training, fresh data should be assigned a higher weight for predicting user's future action.
We assign a weight $w_{new}$ to each new interaction (line~4 of Algorithm~\ref{alg:online}) as a tunable parameter.  Later in Section~\ref{ss:online_eval}, we investigate how the setting of this parameter impacts online learning performance.

\textbf{Time Complexity}. The incremental update for a new interaction $(u,i)$ can be done in $O(K^2 + (|\R_u|+|\R_i|)K)$ time. It is worth noting that the cost depends on the number of observed interactions for $u$ and $i$, while being independent with number of total interactions, users and items. This localized complexity make the online learning algorithm suitable to deployment in industrial use,
as the complex software stack that deals with data dependencies \cite{MLGoogle:2014} can be avoided.

\begin{algorithm}[t]
	\caption{Online Incremental Updates for eALS.\label{alg:online}}
	\KwIn{$\hat{\textbf{P}}$, $\hat{\textbf{Q}}$, new interaction $(u,i)$ and its weight $w_{new}$}
	\KwOut{Refreshed parameters $\textbf{P}$ and $\textbf{Q}$;}
	$\textbf{P}\leftarrow\hat{\textbf{P}};\quad \textbf{Q}\leftarrow\hat{\textbf{Q}};$ \;
	\textbf{If} $u$ is a new user \textbf{do}\quad Randomly initialize $\textbf{p}_u$ \;
	\textbf{If} $i$ is a new item \textbf{do}\quad Randomly initialize $\textbf{q}_i$ \;
	$r_{ui}\leftarrow 1;\quad \hat{r}_{ui}\leftarrow Eq.~(\ref{eq:mf});\quad w_{ui}\leftarrow w_{new}$\; 
	\While{Stopping criteria is not met} {
		\tcp{Line 6-10 of Algorithm 1}
		update\_user($u$); \Comment*[r]{$O(K^2 + |\R_u|K)$}
		update\_cache($u$, $\textbf{S}^p$); \Comment*[r]{$O(K^2)$}
		\tcp{Line 14-18 of Algorithm 1}
		update\_item($i$); \Comment*[r]{$O(K^2 + |\R_i|K)$}
		update\_cache($i$, $\textbf{S}^q$); \Comment*[r]{$O(K^2)$}
	}
	\Return{$\textbf{P}$ and $\textbf{Q}$}
\end{algorithm}

\section{Experiments}
\label{sec:experiments}
We begin by introducing the experimental settings. Then we perform an empirical study with the traditional offline protocol, followed by a more realistic online protocol. 


\subsection{Experimental Settings}
\textbf{Datasets}. 
We evaluate on two publicly accessible datasets: Yelp\footnote{We used the Yelp Challenge dataset downloaded on October 2015 that contained 1.6 million reviews: \scriptsize{\url{http://www.yelp.com/dataset_challenge}}} and Amazon Movies\footnote{\scriptsize{\url{http://snap.stanford.edu/data/web-Amazon-links.html}}}. 
We transform the review dataset into implicit data, where each entry is marked as 0/1 indicating whether the user reviewed the item. 
Since the high sparsity of the original datasets makes it difficult to evaluate recommendation algorithms (\eg over half users have only one review), we follow the common practice \cite{Rendle:2009:BPR} to filter out users and items with less than 10 interactions. Table~\ref{tab:dataset} summarizes the statistics of the filtered datasets. \\\vspace{-5pt}

\begin{table}
\begin{center}
\caption{\textbf{Statistics of the evaluation datasets.}}
\vspace{-8pt}
\small
\label{tab:dataset}
    \begin{tabular}{ | l | c | c | c | c | }
    \hline
    \textbf{Dataset} & \textbf{Review\#} & \textbf{Item\#} & \textbf{User\#}  & \textbf{Sparsity} \\ \hline
    Yelp	& 731,671	& 25,815 & 25,677  & 99.89\% \\ \hline
    Amazon	& 5,020,705	& 75,389 & 117,176  & 99.94 \% \\ \hline
    \end{tabular}
    \vspace{-20pt}
\end{center}
\end{table}

\textbf{Methodology}. We evaluate using two protocols: 

- \textbf{Offline Protocol}. We adopt the \textit{leave-one-out} evaluation, where the latest interaction of each user is held out for prediction and the models are trained on the remaining data. Although it is a widely used evaluation protocol in the literature~\cite{He:2015, Rendle:2009:BPR}, we point out that it is an artificial split that does not correspond to the real recommendation scenario. In addition, the new users problem is averted in this evaluation, as each test user has a training history. Thus this protocol only evaluates an algorithm's capability in providing one-shot recommendation for existing users by leveraging the static history data.

- \textbf{Online Protocol}. To create a more realistic recommendation scenario, we simulate the dynamic data stream. We first sort all interactions in chronological order, training models on the first 90\% of the interactions and holding out the last 10\% for testing. 
In the testing phase, given a test interaction (\ie a user--item pair) from the hold-out data, the model first recommends a ranked list of items to the user; the performance is judged based on the ranked list. 
Then the test interaction is fed into the model for an incremental update. 
Note that with the global split by time, 14\% and 57\% test interactions are from new users for the Yelp and Amazon dataset, respectively. 
Overall this protocol evaluates an online learning algorithm's effectiveness in digesting the dynamic new data.

To assess the ranked list with the ground-truth (GT) item that user actually consumed, we adopt \textit{Hit Ratio} (HR) and \textit{Normalized Discounted Cumulative Gain}~(NDCG). We truncate the ranked list at 100 for both metrics. 
HR measures whether the ground truth item is present on the ranked list, while NDCG accounts for the position of hit \cite{He:2015}.
We report the score averaged by all test interactions. \\ \vspace{-5pt}

\begin{figure*}[t]
	\centering
	\begin{subfigure}[b]{0.25\textwidth}
		\centering
		\includegraphics[width=\textwidth]{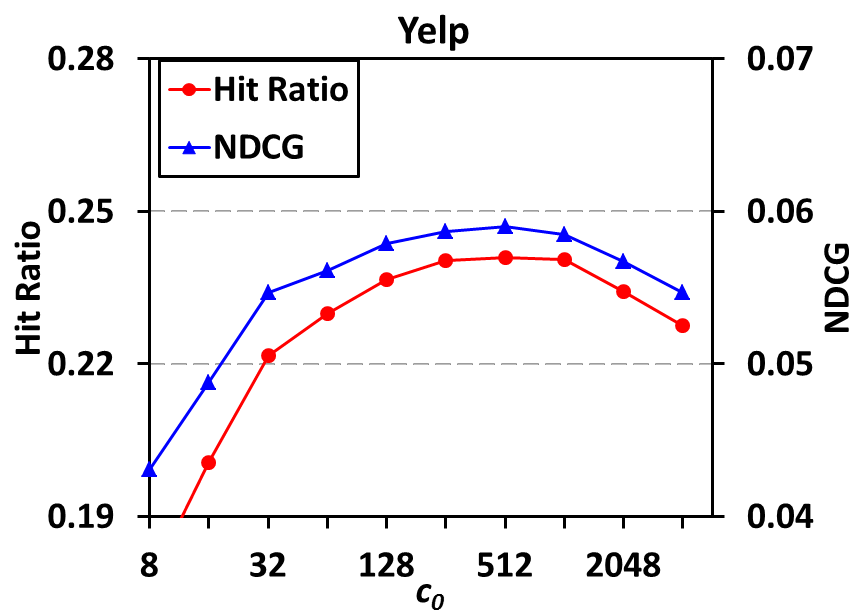}
		\vspace{-15pt}
		\caption{eALS vs. $c_0\ (\alpha=0)$}
		\label{fig:yelp_w0}
	\end{subfigure} \hspace{-7pt}
	\begin{subfigure}[b]{0.25\textwidth}
		\centering
		\includegraphics[width=\textwidth]{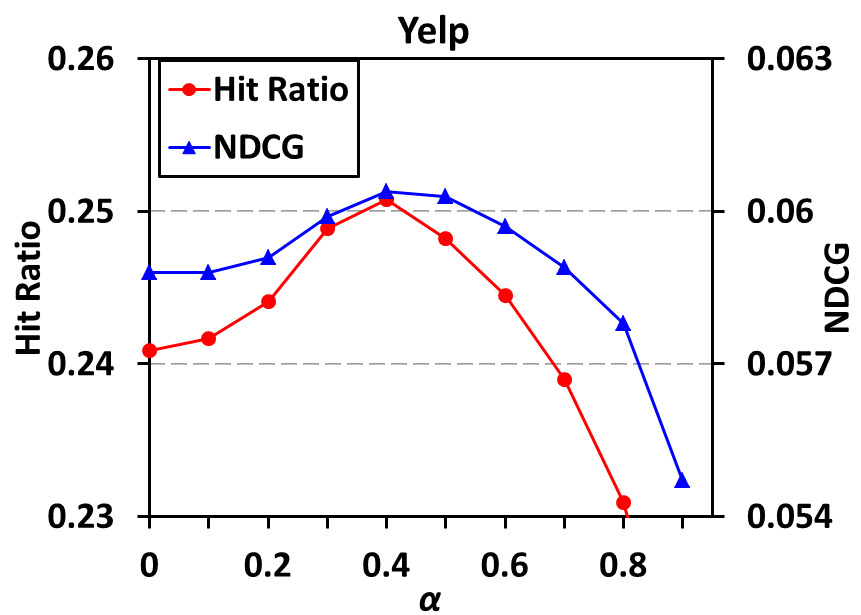}
		\vspace{-15pt}
		\caption{eALS vs. $\alpha\ (c_0=512)$}
		\label{fig:yelp_alpha}
	\end{subfigure} \hspace{-7pt}
	\begin{subfigure}[b]{0.25\textwidth}
		\centering
		\includegraphics[width=\textwidth]{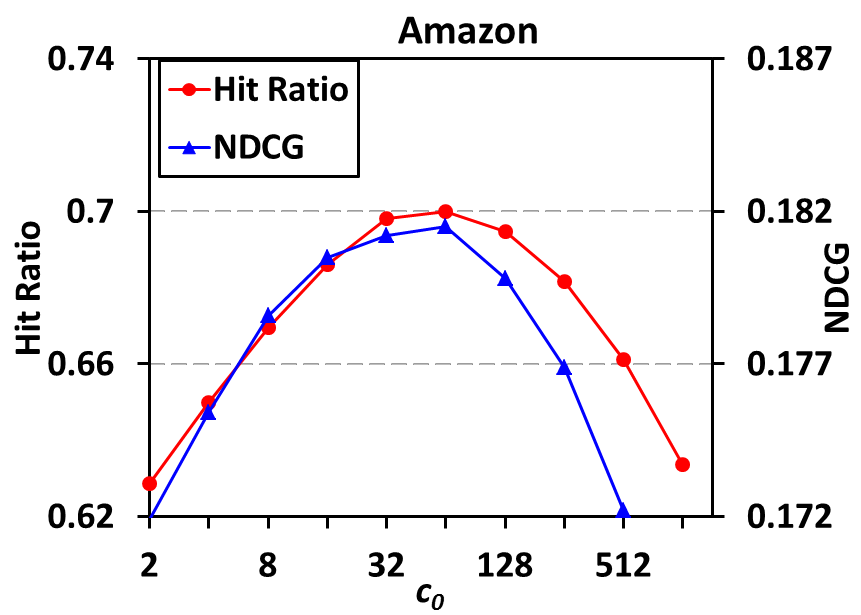}
		\vspace{-15pt}
		\caption{eALS vs. $c_0\ (\alpha=0)$}
		\label{fig:amazon_w0}
	\end{subfigure} \hspace{-7pt}
	\begin{subfigure}[b]{0.25\textwidth}
		\centering
		\includegraphics[width=\textwidth]{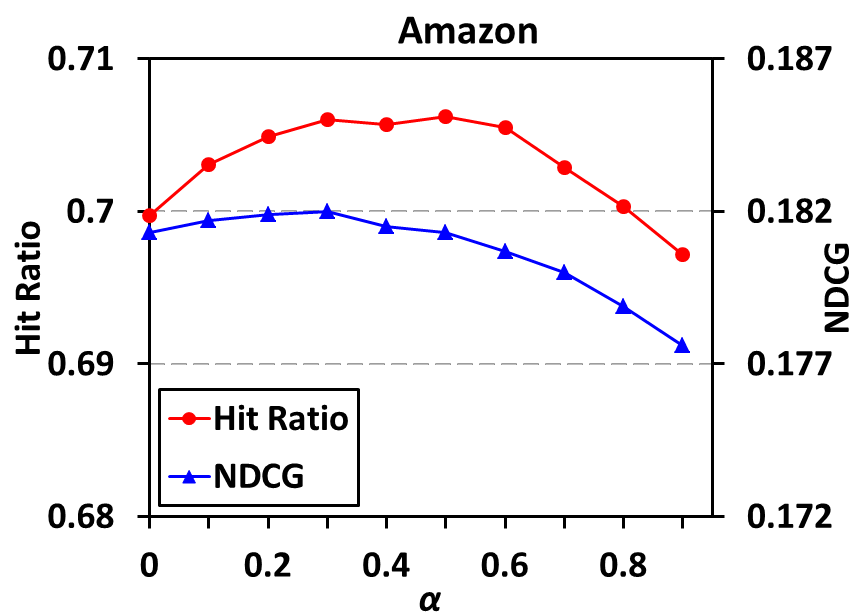}
		\vspace{-15pt}
		\caption{eALS vs. $\alpha\ (c_0=64)$}
		\label{fig:amazon_alpha}
	\end{subfigure} \hspace{-7pt}
	\vspace{-5pt}
	\caption{Impact of weighting parameters $c_0$ and $\alpha$ on eALS's performance evaluated by offline protocol.}
	\vspace{-10pt}
	\label{fig:weight}
\end{figure*}

\begin{figure*}[t]
	\centering
	\begin{subfigure}[b]{0.23\textwidth}
		\centering
		\includegraphics[width=\textwidth]{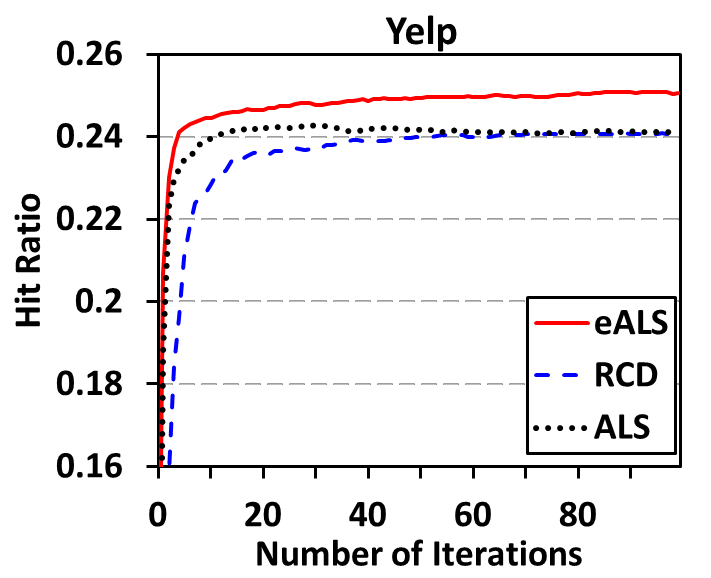}
		\vspace{-10pt}
		\caption{Iterations vs. HR}
		\label{fig:yelp_mf_hr}
	\end{subfigure} 
	\begin{subfigure}[b]{0.23\textwidth}
		\centering
		\includegraphics[width=\textwidth]{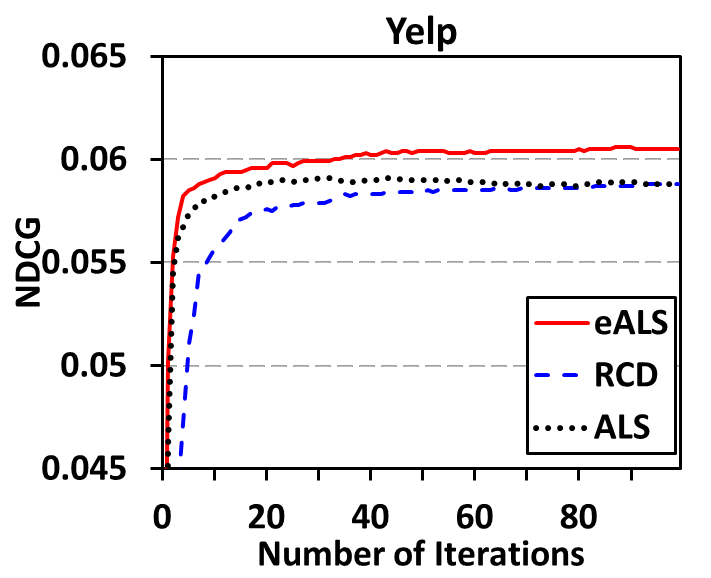}
		\vspace{-10pt}
		\caption{Iterations vs. NDCG}
		\label{fig:yelp_mf_ndcg}
	\end{subfigure} 
	\begin{subfigure}[b]{0.23\textwidth}
		\centering
		\includegraphics[width=\textwidth]{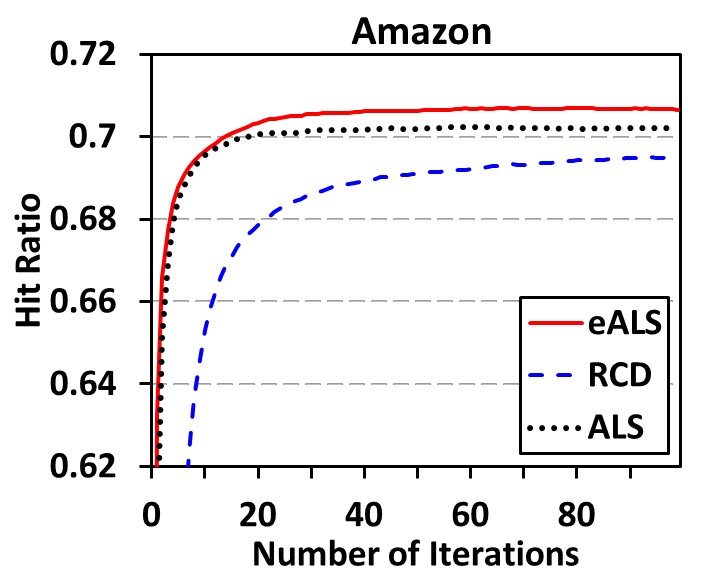}
		\vspace{-10pt}
		\caption{Iterations vs. HR}
		\label{fig:amazon_mf_hr}
	\end{subfigure} 
	\begin{subfigure}[b]{0.23\textwidth}
		\centering
		\includegraphics[width=\textwidth]{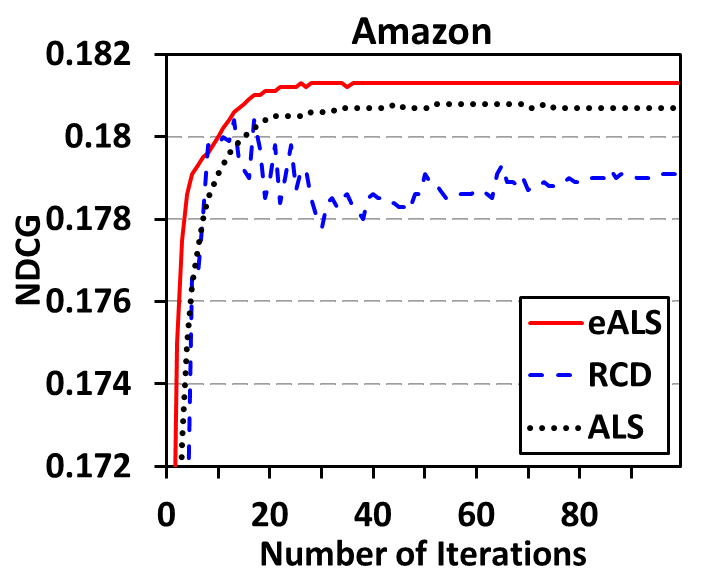}
		\vspace{-10pt}
		\caption{Iterations vs. NDCG}
		\label{fig:amazon_mf_ndcg}
	\end{subfigure} 
	\vspace{-5pt}
	\caption{Prediction accuracy of three whole-data based MF methods in each iteration ($K=128$).}
	\vspace{-10pt}
	\label{fig:mf_convergence}
\end{figure*}

\textbf{Baselines}. We compare with the following methods:

- \textbf{ALS}~\cite{Hu:2008}. This is the conventional ALS method that optimizes the whole-data based MF.
Due to the high time complexity, this method is infeasible in a real-time dynamic updating scenario, so we only evaluate it with the offline protocol.

- \textbf{RCD}~\cite{devooght2015dynamic}. This is the state-of-the-art implicit MF method that has the same time complexity with eALS and is suitable for online learning. 
For the line search parameters, we use the suggested values in the authors' implementation\footnote{\scriptsize{\url{https://github.com/rdevooght/MF-with-prior-and-updates}}}.

- \textbf{BPR}~\cite{Rendle:2009:BPR}. This is a sample-based method that optimizes the pair-wise ranking between the positive and negative samples. 
It learns by SGD, which can be adjusted to online incremental learning by \cite{Rendle:2008}. We use a fixed learning rate, varying it and reporting the best performance. \\\vspace{-5pt}

\textbf{Parameter Settings.}
For the weight of observed interactions, we set it uniformly as 1, a default setting by previous works~\cite{devooght2015dynamic,Rendle:2009:BPR}. 
For regularization, we set $\lambda$ as 0.01 for all methods for a fair comparison. 
All methods are implemented in Java and running on the same machine (Intel Xeon 2.67GHz CPU and 24GB RAM) in a single-thread for a fair comparison on efficiency. As the findings are consistent across the number of factors $K$, without any particular outlier, we only show the results of $K=128$, a relatively large number that maintains good accuracy. 

\subsection{Offline Protocol}
We first study how does the weighting scheme on missing data impact eALS's performance. Then we compare with the whole-data based implicit MF methods ALS and RCD, as well as the sample-based ranking method BPR. 

\subsubsection{Weight of Missing Data}
In Section \ref{ss:popularity}, we propose an item popularity-aware weighting strategy, which has two parameters: $c_0$ determines the overall weight of missing data and $\alpha$ controls the weight distribution. 
First, we set a uniform weight distribution (\ie $\alpha = 0$), varying $c_0$ to study how does the weight of missing data impact the performance.
For Yelp (Figure~\ref{fig:yelp_w0}), the peak performance is achieved when $c_0$ is around 512, corresponding to that the weight of each zero entry is $0.02$ ($w_0=c_0/N$); similarly for Amazon (Figure~\ref{fig:amazon_w0}), the optimal $c_0$ is around 64, corresponding to $w_0=0.0001$. When $c_0$ becomes smaller (where $w_0$ is close to 0), the performance degrades significantly. 
This highlights the necessity of accounting for the missing data when modeling implicit feedback for item recommendation. Moreover, when $c_0$ is set too large, the performance also suffers.
Based on this observation, we believe the traditional SVD technique~\cite{Cremonesi:2010} that treats all entries equally weighted will be suboptimal here. 

Then, we set $c_0$ to the best value (in the case of $\alpha=0$), varying $\alpha$ to check the performance change.  As can be seen from Figure~\ref{fig:yelp_alpha} and \ref{fig:amazon_alpha}, the performance of eALS is gradually improved with the increase of $\alpha$, and the best result is reached around 0.4. 
We further conducted the one-sample paired $t$-test, verifying that the improvements are statistically significant 
($p$-value < 0.01) for both metrics on the two datasets. 
This indicates the effectiveness of our popularity-biased weighting strategy. 
Moreover, when $\alpha$ is larger than 0.5, the performance starts to drop significantly.
This reveals the drawback of over-weighting popular items as negative instances, thus the importance of accounting for less popular items with a proper weight. 

In the following experiments, we fix $c_0$ and $\alpha$ according to the best performance evaluated by HR, \ie $c_0=512, \alpha=0.4$ for Yelp and $c_0=64, \alpha=0.5$ for Amazon. 

\subsubsection{Compare Whole-data based MF Methods}
\label{ss:mf_learners}
We performed the same grid search of $w_0$ for RCD and ALS and reported the best performance. 

\textbf{Convergence}. Figure~\ref{fig:mf_convergence} shows the prediction accuracy with respect to number of iterations.
First, we see that eALS achieves the best performance after converge. 
All improvements are statistically significant evidenced by the one-sample paired $t$-test ($p<0.01$).
We believe the benefits mainly come from the popularity-aware objective function, as both ALS and RCD apply a uniform weighting on the unknowns. 
Second, eALS and ALS converge faster than RCD. We think the reason is that (e)ALS updates a parameter to minimize the objective function of the current status, while RCD updates towards the direction of the negative gradient, which can be suboptimal. On Amazon, RCD shows high but turbulent NDCG in early iterations, while the low hit ratio and later iterations indicate the high NDCG is unstable. 
Finally, we point out that in optimizing the same objective function, ALS outperforms RCD in most cases, demonstrating the advantage of ALS over the gradient descent learner. 

\begin{figure}
	\centering
	\begin{subfigure}[b]{0.23\textwidth}
		\centering
		\includegraphics[width=\textwidth]{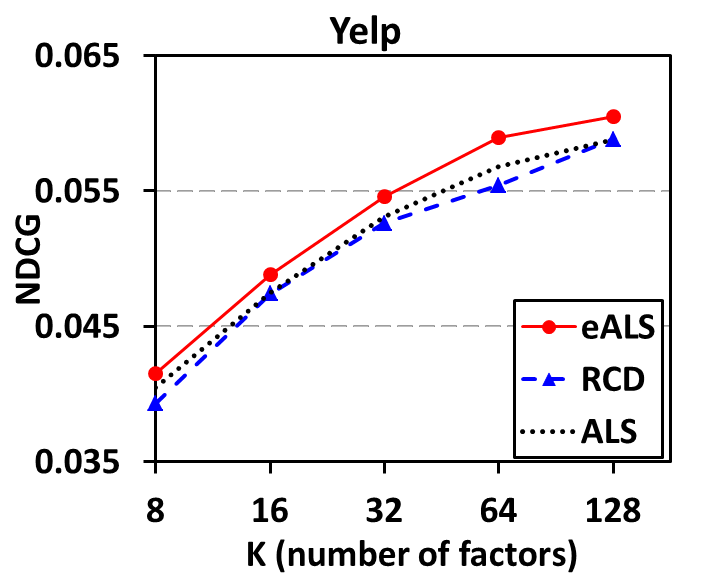}
		\vspace{-15pt}
		\label{fig:yelp_k_ndcg}
	\end{subfigure} 
	\begin{subfigure}[b]{0.23\textwidth}
		\centering
		\includegraphics[width=\textwidth]{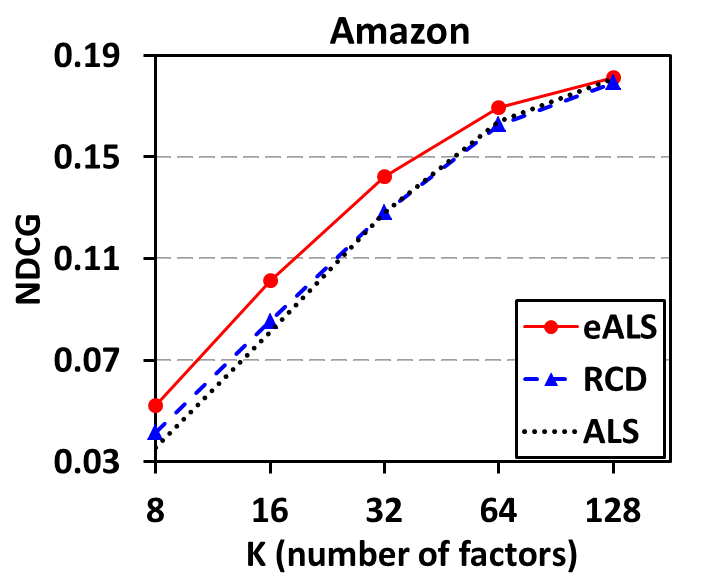}
		\vspace{-15pt}
		\label{fig:amazon_k_ndcg}
	\end{subfigure} 
	\caption{NDCG of whole-data based MF across $K$.}
	\vspace{-10pt}
	\label{fig:mf_k}
\end{figure}

\begin{figure*}[t]
	\centering
	\begin{subfigure}[b]{0.25\textwidth}
		\centering
		\includegraphics[width=\textwidth]{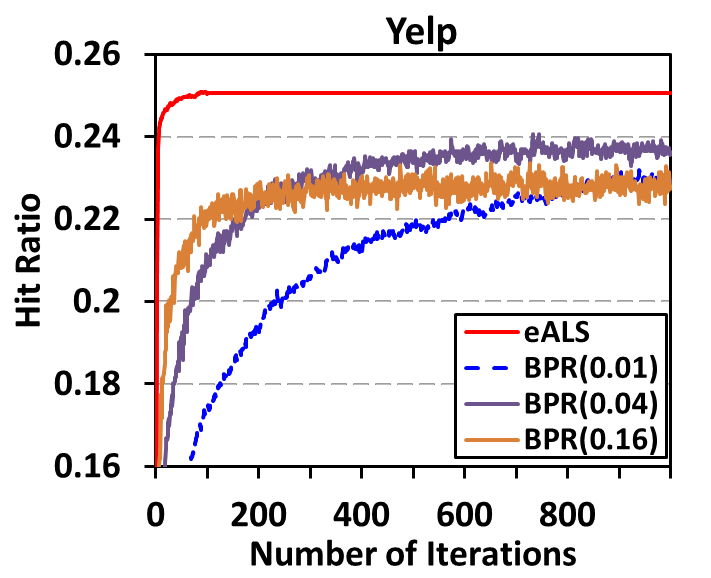}
		\vspace{-15pt}
		\caption{Iterations vs. HR}
		\label{fig:yelp_bpr_hr}
	\end{subfigure} \hspace{-7pt}
	\begin{subfigure}[b]{0.25\textwidth}
		\centering
		\includegraphics[width=\textwidth]{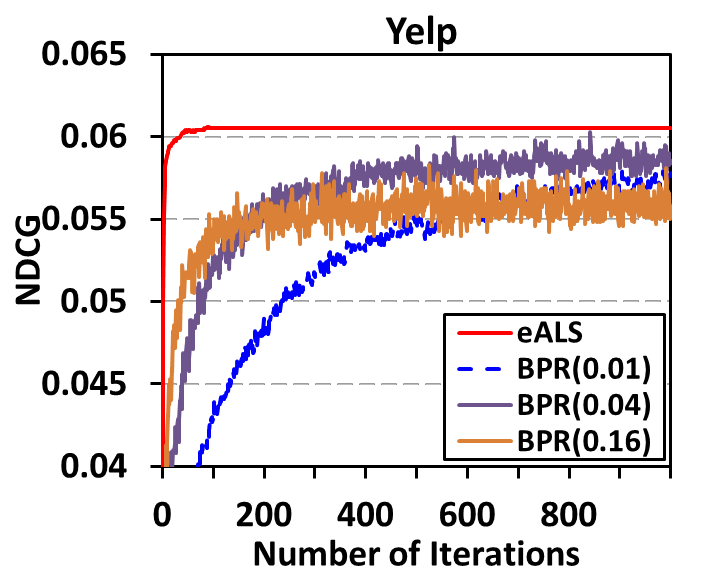}
		\vspace{-15pt}
		\caption{Iterations vs. NDCG}
		\label{fig:yelp_bpr_ndcg}
	\end{subfigure} \hspace{-7pt}
	\begin{subfigure}[b]{0.25\textwidth}
		\centering
		\includegraphics[width=\textwidth]{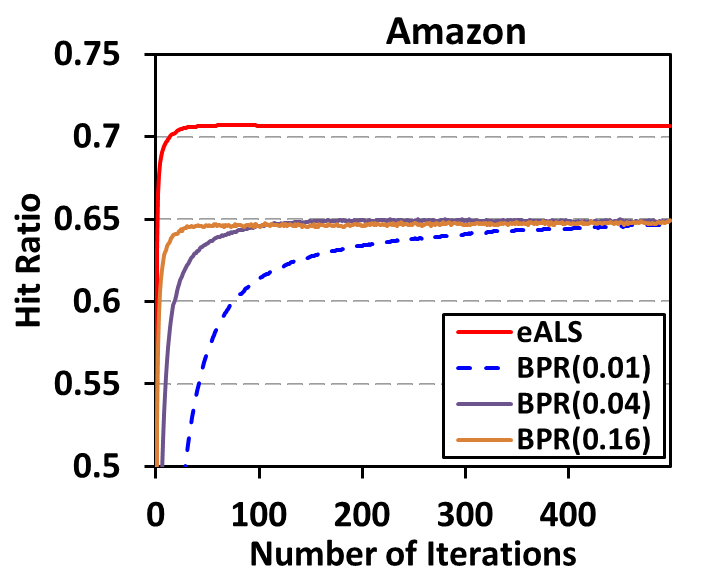}
		\vspace{-15pt}
		\caption{Iterations vs. HR}
		\label{fig:amazon_bpr_hr}
	\end{subfigure} \hspace{-7pt}
	\begin{subfigure}[b]{0.25\textwidth}
		\centering
		\includegraphics[width=\textwidth]{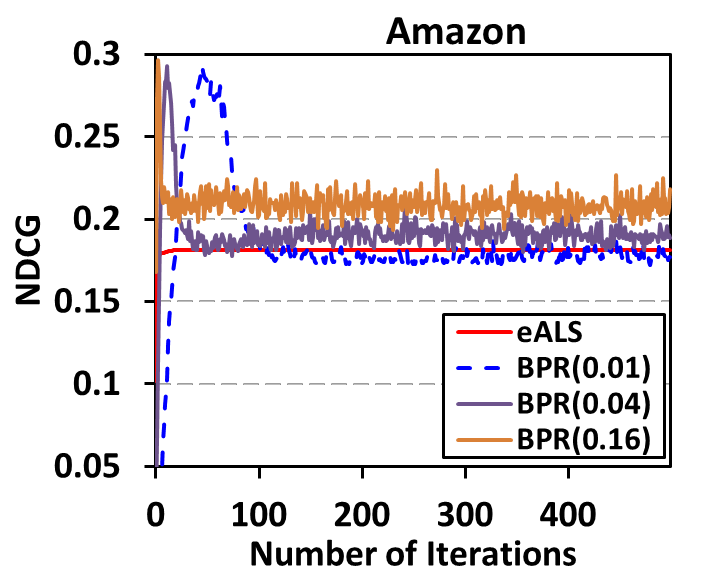}
		\vspace{-15pt}
		\caption{Iterations vs. NDCG}
		\label{fig:amazon_bpr_ndcg}
	\end{subfigure} \hspace{-7pt}
	\vspace{-5pt}
	\caption{Accuracy and convergence comparison of eALS with BPR of different learning rate.}
	\vspace{-10pt}
	\label{fig:bpr}
\end{figure*}

\textbf{Accuracy vs. Number of Factors}. Figure~\ref{fig:mf_k} shows the prediction accuracy with varying number of factors $K$. We only show the evaluation by NDCG as HR admits the same trend. 
First, eALS consistently outperforms ALS and RCD across $K$, demonstrating the effectiveness of our eALS method (\textit{n.b.} although the three methods seem to perform on par for Amazon at $K=128$, their difference can be clearly seen in  Figure~\ref{fig:amazon_mf_ndcg}).
Second, all methods can be improved significantly with a larger $K$. Although a large $K$ might have the risk of overfitting, it can increase model's representation ability thus better prediction. Especially for large datasets that can have millions of users and billions of interactions, a large $K$ is particularly important for the accuracy of MF methods. 

\textbf{Efficiency.} 
Analytically, ALS's time complexity is $O((M+N)K^3 + |\R|K^2)$, while eALS and RCD are $K$ times faster. 
To compare their efficiency empirically, we show the actual training time per iteration in Table~\ref{tab:efficency}. 

\vspace{-5pt}
\begin{table}[h]
	\begin{center}
		\caption{Training time per iteration of different whole-based MF methods with varying $K$.}
		\small
		\vspace{-8pt}
		\label{tab:efficency}
		\begin{tabular}{ | c | c | c | c | c | c | c |}
			\hline
			& \multicolumn{3}{c}{\textbf{Yelp}} & \multicolumn{3}{|c|}{\textbf{Amazon}} \\ \hline 
			\textbf{K} & \textbf{eALS} & \textbf{RCD} & \textbf{ALS} & \textbf{eALS} & \textbf{RCD} & \textbf{ALS} \\ \hline
			\textbf{32}	& 1s & 1s & 10s & 9s & 10s & 74s \\ \hline
			\textbf{64}	& 4s & 3s & 46s & 23s & 17s & 4.8m \\ \hline
			\textbf{128}	& 13s & 10s & 221s & 72s & 42s & 21m \\ \hline
			\textbf{256}	& 1m& 0.9m& 23m & 4m & 2.8m & 2h \\ \hline
			\textbf{512}	& 2m& 2m & 2.5h & 12m & 9m & 11.6h \\ \hline
			\textbf{1024}	& 7m& 11m & 25.4h & 54m & 48m & 74h  \\ \hline
		\end{tabular}
	\end{center}
	\vspace{-8pt}
	\scriptsize{\quad$s$, $m$, and $h$ denote seconds, minutes and hours, respectively.}
	\vspace{-5pt}
\end{table}

As can be seen, with the increase of $K$, ALS takes much longer time than eALS and RCD. Specifically, when $K$ is 512, ALS requires 11.6 hours for one iteration on Amazon, while eALS only takes 12 minutes. 
Although eALS does not empirically show $K$ times faster than ALS due to the more efficient matrix inversion implementation (we used the fastest known algorithm \cite{Coppersmith:1990} with time complexity around $O(K^{2.376})$), the speed-up is already very significant.
Moreover, as RCD and eALS have the same analytical time complexity, their actual running time are in the same magnitude; the minor difference can be caused by some implementation details, such as the data structures and caches used. 

\subsubsection{eALS vs. BPR (sample-based)}


Figure~\ref{fig:bpr} plots the performance of BPR with different learning rates\footnote{We have also tried other intermediate values of learning rates, and the findings are consistent. Thus, to make the figure more clear, we only show three selected values.} in each iteration. Note that we only run eALS for 100 iterations, which are enough for eALS to converge. First, it is clear that BPR's performance is subjected to the choice of learning rate --- Figure~\ref{fig:yelp_bpr_hr} and \ref{fig:yelp_bpr_ndcg} show that a higher learning rate leads to a faster convergence, while the final accuracy may be suffered. 
Second, we see that eALS significantly outperforms BPR on the Yelp dataset evaluated by both measures ($p<0.001$). For Amazon, eALS obtains a much higher hit ratio but a lower NDCG score, indicating that most hits occur at a relatively low ranks for eALS. Comparing with the performance of other whole-based MF methods ALS and RCD (Figure~\ref{fig:mf_convergence}), we draw the conclusion that BPR is a weak performer in terms of the prediction recall, while being a strong performer in terms of the precision at top ranks. 
We think BPR's strength in ranking top items is due to its optimization objective, which is a pair-wise ranking function tailored for ranking correct item high. In contrast, the regression-based objective is not directly optimized for ranking; instead, by account for all missing data in regression, it better predicts user's preference on unconsumed items, leading to a better recall. This is consistent with \cite{Cremonesi:2010}'s finding in evaluating top-K recommendation. 

We notice that BPR shows unusual NDCG spike in early iterations on the Amazon dataset, however the performance is unstable and goes down with more iterations. 
The same phenomenon was also observed for another gradient descent method RCD on the same dataset (see Figure~\ref{fig:amazon_mf_ndcg}).
We hypothesize that it might be caused by some regularities in the data. For example, we find some Amazon users review on a movie multiple times\footnote{Due to user's repeat consumption behaviours, we do not exclude training items when generating recommend list. 
}. In early iterations, BPR ranks these repeated items high, leading to a high but unstable NDCG score. There might be other reasons responsible for this, and we do not further explore here. \\\vspace{-5pt}


\subsection{Online Protocol}
\label{ss:online_eval}
In the evaluation of online protocol, we hold out the latest 10\% interactions as the test set, training all methods on the remaining 90\% data with the best parameter settings evidenced by the offline evaluation. We first study the number of online iterations required for eALS to converge. Then we show how does the weight of new interactions impact the performance. Lastly, we compare with dynamic MF methods RCD and BPR in the online learning scenario. 

\begin{figure*}[t]
	\centering
	\begin{subfigure}[b]{0.25\textwidth}
		\centering
		\includegraphics[width=\textwidth]{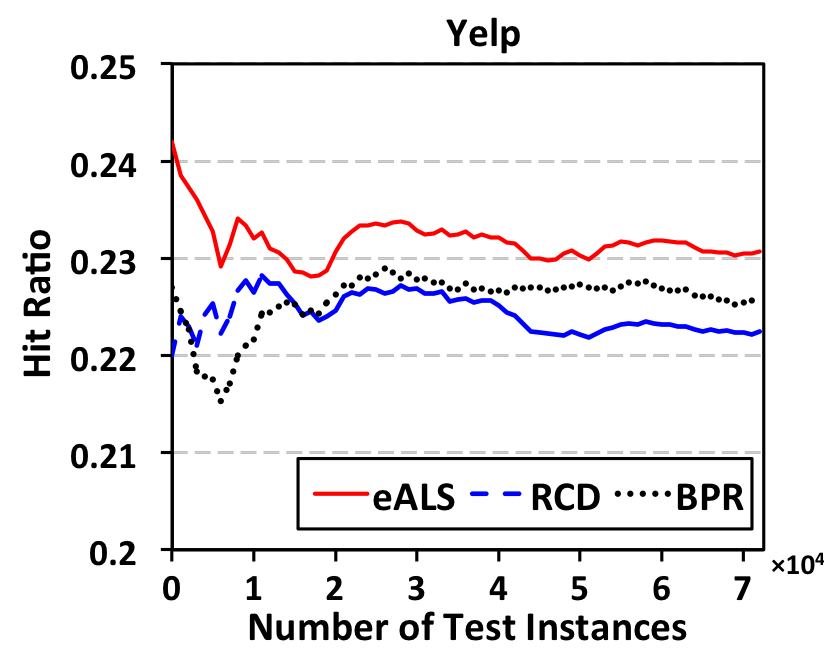}
		\vspace{-15pt}
		\caption{Test \# vs. HR}
		\label{fig:yelp_online_hr}
	\end{subfigure} \hspace{-7pt}
	\begin{subfigure}[b]{0.25\textwidth}
		\centering
		\includegraphics[width=\textwidth]{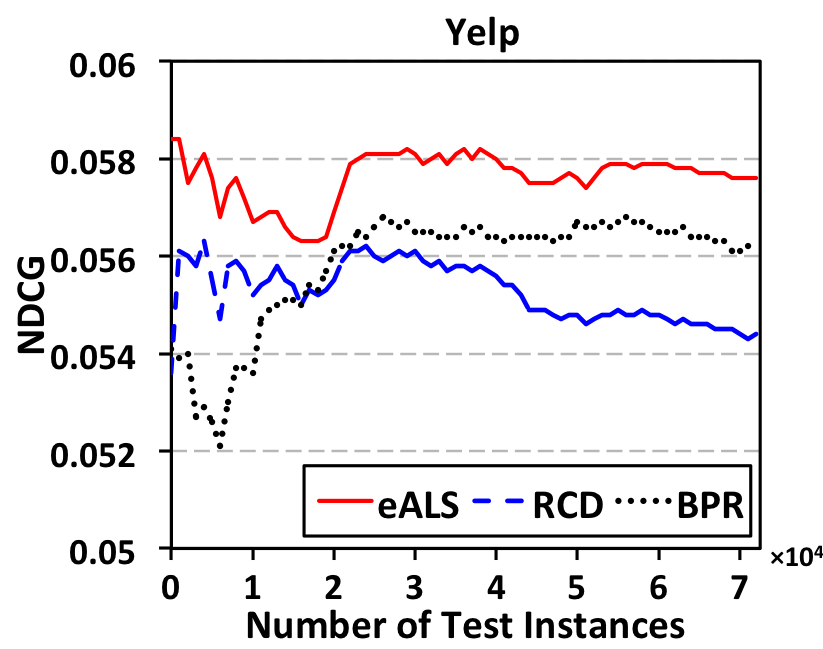}
		\vspace{-15pt}
		\caption{Test \# vs. NDCG}
		\label{fig:yelp_online_ndcg}
	\end{subfigure} \hspace{-7pt}
	\begin{subfigure}[b]{0.25\textwidth}
		\centering
		\includegraphics[width=\textwidth]{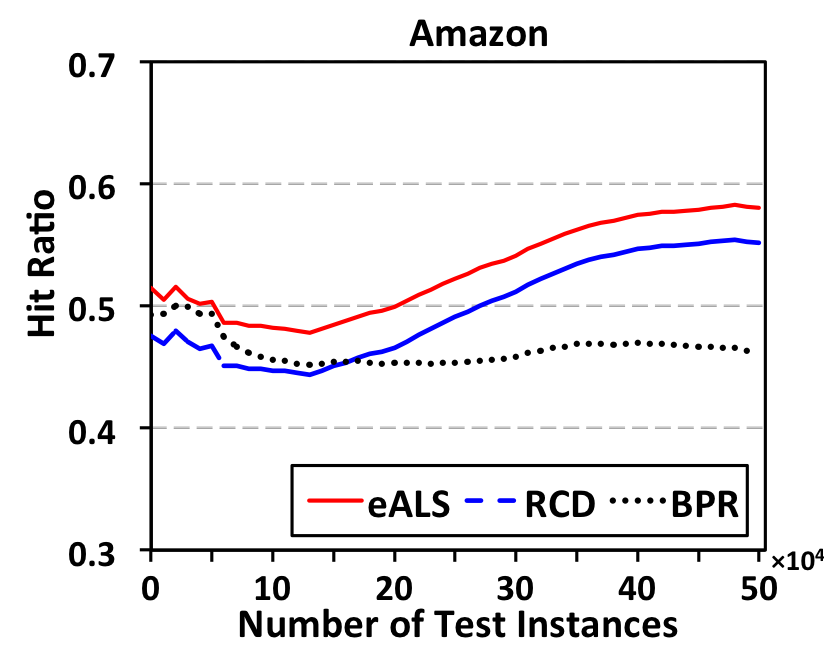}
		\vspace{-15pt}
		\caption{Test \# vs. HR}
		\label{fig:amazon_online_hr}
	\end{subfigure} \hspace{-7pt}
	\begin{subfigure}[b]{0.25\textwidth}
		\centering
		\includegraphics[width=\textwidth]{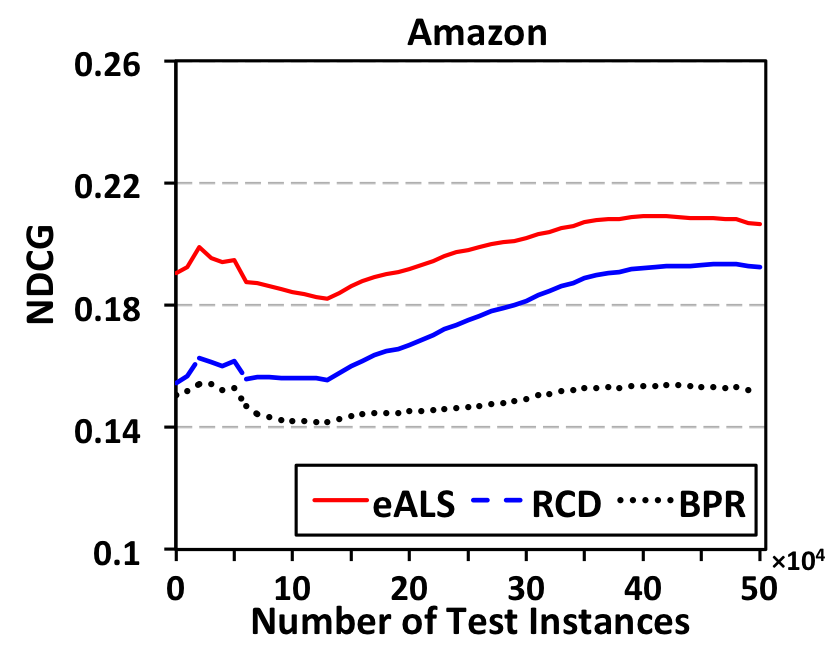}
		\vspace{-15pt}
		\caption{Test \# vs. NDCG}
		\label{fig:amazon_online_ndcg}
	\end{subfigure} \hspace{-7pt}
	\vspace{-5pt}
	\caption{Performance evolution of eALS and other dynamic MF methods in online learning.}
	\vspace{-10pt}
	\label{fig:performance_online}
\end{figure*}

\subsubsection{Number of Online Iterations}

\begin{figure}[h]
	\centering
	\begin{subfigure}[b]{0.24\textwidth}
		\centering
		\includegraphics[width=\textwidth]{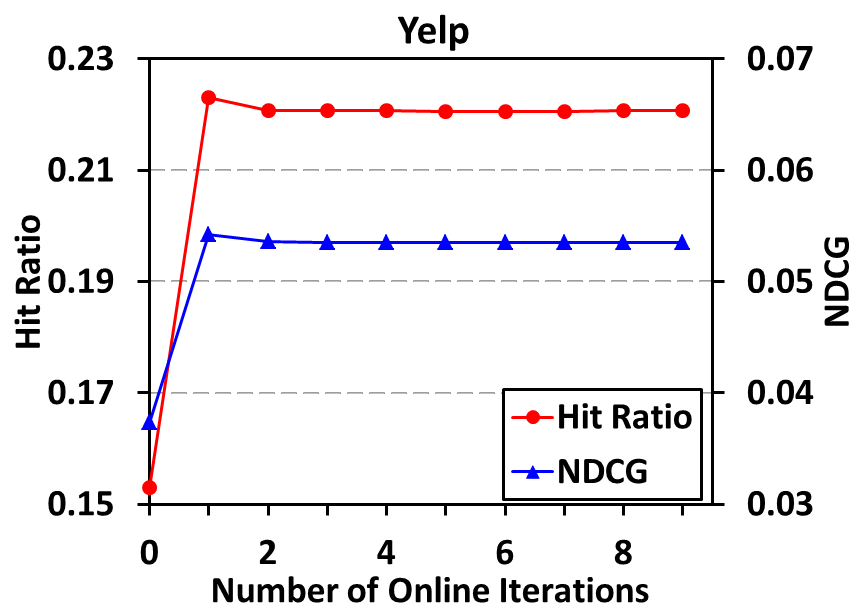}
		\vspace{-15pt}
	\end{subfigure} \hspace{-7pt}
	\begin{subfigure}[b]{0.24\textwidth}
		\centering
		\includegraphics[width=\textwidth]{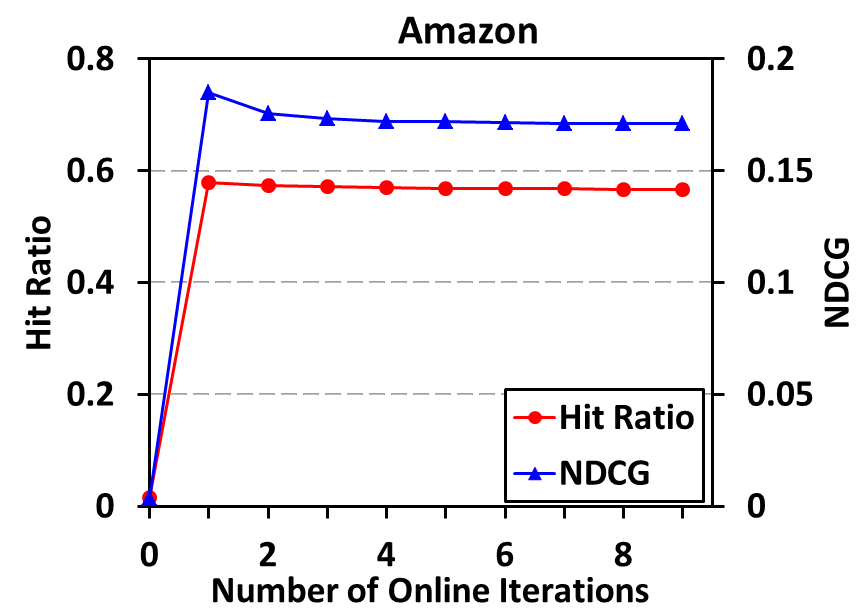}
		\vspace{-15pt}
	\end{subfigure} \hspace{-7pt}
	\caption{Impact of online iterations on eALS.}
	\label{fig:online_iter}
	\vspace{-5pt}
\end{figure}

Figure~\ref{fig:online_iter} shows how does eALS's accuracy change with number of online iterations. Results at the 0-th iteration benchmark the performance of the offline trained model, as no incremental update is performed. First, we can see that the offline trained model performs very poorly, highlighting the importance of refreshing recommender model for an online system with dynamic data. Second, we find most performance gain comes from the first iteration, and more iterations do not further improve. 
This is due to the fact that only the local features regarding to the new interaction are updated, and one eALS step on a latent factor can find the optimal solution with others fixed. Thus, one iteration is enough for eALS to learn from a new interaction incrementally, making eALS very efficient for learning online. 
\vspace{-5pt}

We have also investigated number of online iterations required for baselines RCD and BPR. RCD shows the same trend that good prediction is obtained in the first iteration. While BPR requires more iterations, usually 5-10 iterations to get a peak performance and more iterations will adversely hurt the performance due to the local over-training. 
 
\subsubsection{Weight of New Interactions}

To evaluate how does the weight of new interactions effect the online learning algorithms, we also apply the same weight $w_{new}$ on RCD. Note that the original RCD paper~\cite{devooght2015dynamic} does not consider the weight of interactions; we encode $w_{new}$ the same way with eALS, revising the RCD learner to optimize the weighted regression function.
\vspace{-5pt}
\begin{figure}[h]
	\centering
	\begin{subfigure}[b]{0.22\textwidth}
		\centering
		\includegraphics[width=\textwidth]{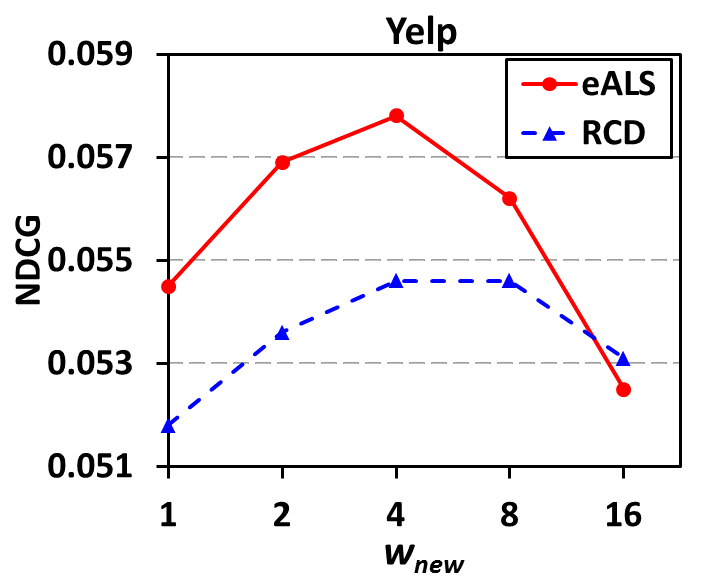}
		\vspace{-15pt}
		\label{fig:yelp_w_ndcg}
	\end{subfigure} 
	\begin{subfigure}[b]{0.22\textwidth}
		\centering
		\includegraphics[width=\textwidth]{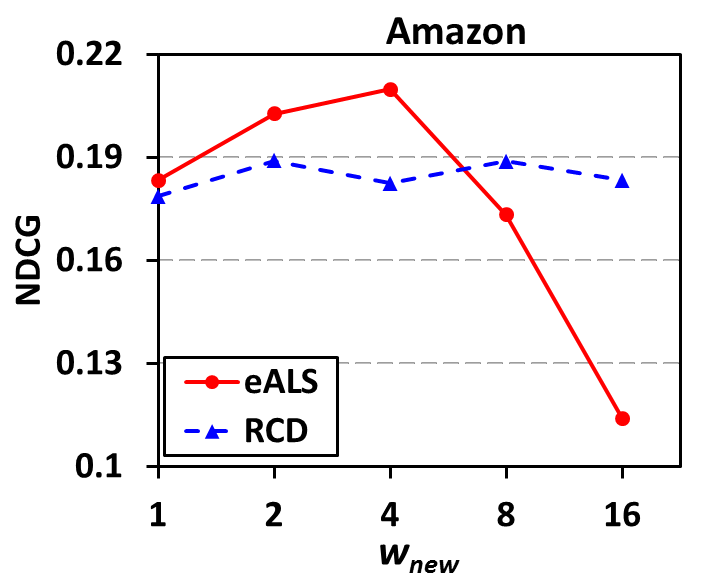}
		\vspace{-15pt}
		\label{fig:amazon_w_ndcg}
	\end{subfigure} 
	\vspace{-5pt}
	\caption{Impact of $w_{new}$ on eALS and RCD in online learning evaluated by NDCG.}
	\vspace{-5pt}
	\label{fig:w_new}
\end{figure}

Figure~\ref{fig:w_new} shows the performance evaluated by NDCG (results of HR show the same trend thus omitted for space). Setting $w_{new}$ to 1 signifies that new interaction is assigned a same weight with the old training interaction. As expected, with a modest increasing on $w_{new}$, the prediction of both models is gradually improved, demonstrating the usefulness of strengthening user's short-term interest.
The peak performance is obtained around 4, where eALS shows better prediction than RCD. 
Overly increasing $w_{new}$ will adversely hurt the performance, admitting the utility of user's historical data used in offline training. Overall, this experiment indicates the importance of balancing user's short-term and long-term interest for quality recommendation. 

\subsubsection{Performance Comparison}

\begin{figure}[h]
	\centering
	\begin{subfigure}[b]{0.24\textwidth}
		\centering
		\includegraphics[width=\textwidth]{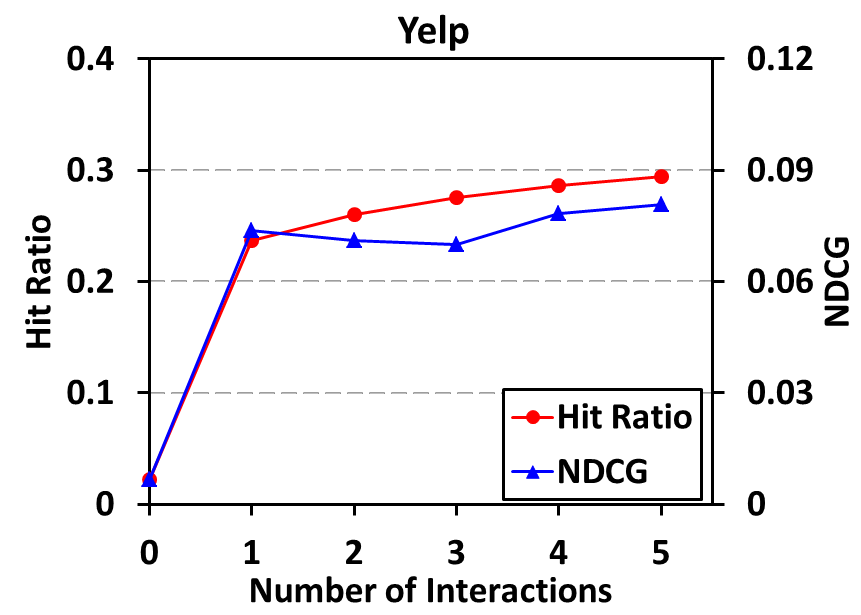}
		\vspace{-15pt}
		\label{fig:yelp_interactions}
	\end{subfigure} \hspace{-7pt}
	\begin{subfigure}[b]{0.24\textwidth}
		\centering
		\includegraphics[width=\textwidth]{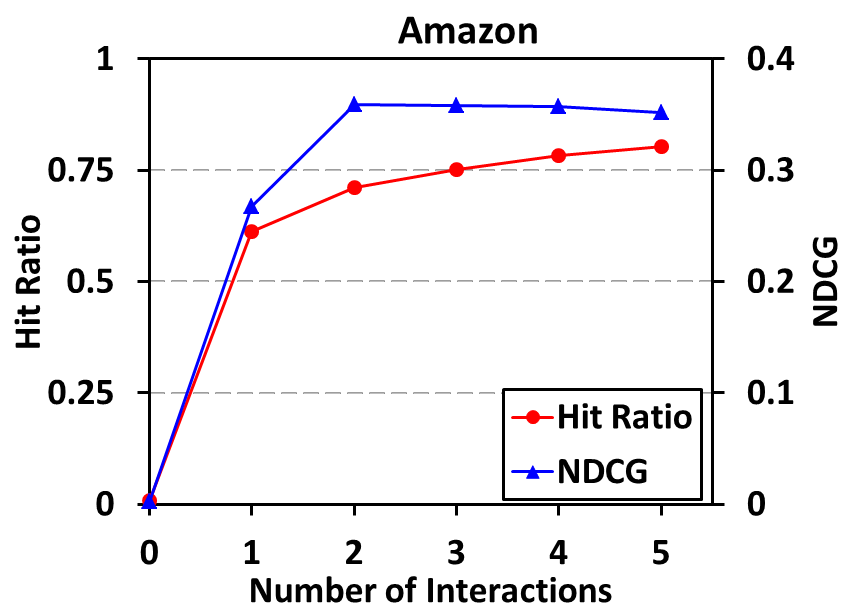}
		\vspace{-15pt}
		\label{fig:amazon_interactions}
	\end{subfigure} \hspace{-7pt}
	\caption{Results breakdown of eALS by \# of past interactions of test user. Note: Interaction \# > 0 denotes the performance for non-cold-start users.}
	\vspace{-5pt}
	\label{fig:interactions}
\end{figure}

With the simulated data stream, we show the performance evolution with respect to number of test instances in Figure \ref{fig:performance_online}. First, eALS consistently outperforms RCD and BPR evidenced by both measures, and 
one-sample paired t-test verifies that all improvements are statistically significant with $p<0.001$. 
BPR betters RCD for Yelp, while underperforms for Amazon. 
Second, we observe the trend that the performance of dynamic learning first decreases, and then increases before becoming stable. This is caused by the new users problem --- when there are few feedback for a user, the model can not personalize the user's preference effectively; with more feedbacks streaming in, the model can adapt itself to improve the preference modeling accordingly. To show this, we further breakdown the results of eALS by number of past interactions
of test user in Figure~\ref{fig:interactions}. 
\vspace{-5pt}

It is clear that when there are no historical feedback for a test user (\ie user cold-start cases), the performance is very poor --- no better than random. 
After the first interaction streams in, the prediction is significantly improved; and with more interactions, the performance is further improved. This highlights the importance of incorporating instantaneous user feedback into the model, especially for cold-start or sparse users that have few history in training.



\section{Conclusion and Future Work}
\label{sec:conclusion}

We study the problem of learning MF models from implicit feedback. 
In contrast to previous work that applied a uniform weight on missing data, 
we propose to weight missing data based on the popularity of items. 
To address the key efficiency challenge in optimization, we develop a new learning algorithm --- eALS --- which effectively learns parameters by performing coordinate descent with memoization. 
For online learning, we devise an incremental update strategy for eALS to adapt dynamic data in real-time. Experiments with both offline and online protocols demonstrate promising results.  Importantly, our work makes MF more practical to use for modeling implicit data, along two dimensions. First, we investigate a new paradigm to deal with missing data which can easily incorporate prior domain knowledge. 
Second, eALS is embarrassingly parallel, making it attractive
for large-scale industrial deployment.

We plan to study the optimal weighting strategy for online data as a way to explore user's short-term interest. 
Along the technical line, we explored the element-wise ALS learner in its basic MF form and solved the efficiency challenge in handling missing data.
To make our method more applicable to real-world settings, we plan to encode side information such as user social contexts~\cite{Geng:2015} and reviews~\cite{He:2015} by extending eALS to more generic models, such as collective factorization~\cite{He:WWW2014} and Factorization machines~\cite{libfm}. In addition, we will study binary coding for MF on implicit data, since a recent advance \cite{DCF:2016} has shown that discrete latent factors are beneficial to collaborative filtering for explicit ratings. 

The strength of eALS can be applied to other domains, owing to the universality of factorizing sparse data matrices.
For example, recent advances in natural language processing~\cite{levy2014neural} have shown the connection between neural word embeddings and MF on the word--context matrix. This bridge nicely motivates several proposals to use MF to learn word embeddings; however, when it comes to handling missing data, they have either ignored~\cite{pennington2014glove} or equally weighted the missing entries, similar to traditional SVD~\cite{levy2014neural}. 
It will be interesting to see whether eALS will also improve these tasks.

\vspace{+5pt}
\noindent {\large \textbf{Acknowledgement}}

\noindent 
The authors would like to thank the additional discussion and help
from Steffen Rendle, Bhargav Kanagal, Immanuel Bayer, Tao Chen, Ming
Gao and Jovian Lin.
\vspace{-10pt}


\bibliographystyle{abbrv}
{\small\bibliography{proc-abbrv}}
\end{document}